%% file: main.tex
\documentclass{article}
\usepackage{spconf,amsmath,graphicx,url,multirow,gensymb}
\usepackage[nolist]{acronym}
\usepackage{xcolor}
\usepackage{booktabs}
\usepackage{subfigure} 

\title{ {SL\MakeLowercase{o}C\MakeLowercase{las}: A Database for Joint Sound Localization and Classification}}

\name{Xinyuan Qian, Bidisha Sharma, Amine El Abridi and Haizhou Li
\thanks{This research is supported by the Agency for Science, Technology and Research (A*STAR) under its AME Programmatic Funding Scheme (Project No. A18A2b0046) and Science and Engineering Research Council, Agency of Science, Technology and Research, Singapore, through the National Robotics Program under Grant No. 192 25 00054.}}
\address{Department of Electrical and Computer Engineering,\\
  National University of Singapore, Singapore}

\begin{document}
%\ninept
%
\include{acro}

\maketitle
\newcommand{\todo}[1]{\color{green} #1}
\newcommand{\etal}{\textit{et al.}}
\newcommand{\eg}{\textit{e.g.}}
\newcommand{\etc}{\textit{e.t.c.}}
\newcommand{\ie}{\textit{i.e.}}
\newcommand{\algs}{\textit{Algorithm}}

\newcommand{\fig}[1]{Figure }
\newcommand{\tab}[1]{Table}
\newcommand{\Sec}[1]{Section }
\newcommand{\eq}[1]{Eq.}
\newcommand{\qian}[1]{{\color{blue}#1}}

\begin{abstract}
In this work, we present the development of a new database, namely Sound Localization and Classification ({\it SLoClas}) corpus, for studying and analyzing  sound localization and classification. The corpus contains a total of 23.27 hours of data recorded using a 4-channel microphone array. 10 classes of sounds are played over a loudspeaker at 1.5 meters distance from the array by varying the \ac{DoA} from 1$^\circ$ to 360$^\circ$ at an interval of $5^\circ$. To facilitate the study of noise robustness,  6  types of outdoor noise are recorded at 4 \ac{DoA}s, using the same devices. Moreover, we propose a baseline method, namely \ac{SLCnet} and present the experimental results and analysis conducted  on the collected SLoClas database. We achieve the accuracy of 95.21\% and 80.01\% for sound localization and classification, respectively. We publicly release this database and the source code for research purpose.
\end{abstract}
\vspace{-0.4cm}
\section{Introduction}\label{sec:intro}
\ac{SLC} refers to estimating the spatial location of a sound source and identifying the type of a sound event through a unified framework. A \ac{SLC} method enables the autonomous robots to determine sound location and detect sound events for navigation and interaction with the surroundings~\cite{takeda2016discriminative,sharma2019multi}. Thus, \ac{SLC} is useful in smart-city and smart-home applications to automatically specify social or human activities, and assist the hearing impaired to visualize and realize sounds~\cite{wang2021gcc,qian2021multi,zhang2021deep,tao2021is}. 
Specifically, a \ac{SLC} method can be divided into two sub-tasks, \ac{DOAE} and \ac{SEC}.

Most traditional \ac{DOAE} parametric approaches rely on \ac{TDOA} estimation~\cite{grobler2017sound, valenzise2007scream}, maximization of \ac{SRP}~\cite{butko2011two}, or acoustic intensity vector analysis~\cite{lopatka2016detection}. 
Whereas the \ac{SEC} methods rely on both low-level and high-level attributes, including spectro-temporal and \ac{MFCC} features~\cite{foggia2015audio} 
where classification-based frameworks such as \ac{GMM}~\cite{valenzise2007scream}, \ac{HMM}~\cite{butko2011two}, and \ac{SVM}~\cite{politis2017comparing}, are mostly adopted. 
% Most of these methods consider the localization and classification as two separate tasks.

Learning-based approaches have the advantages of good generalization under different acoustic environments, where they enable a system to learn the mapping between input features and  outputs.
Recently, deep-learning based approaches have been successfully applied  on \ac{DOAE}~\cite{adavanne2018direction,perotin2019crnn} and \ac{SEC}~\cite{mesaros2019sound}.
Since in real-world applications, a sound event may transmit from a specific direction, it is reasonable to combine \ac{DOAE} and \ac{SEC} with not only estimating the respective associated spatial location, but also detecting the type of sound \cite{cao2019polyphonic}.
Therefore, it is worthy to investigate the influences and potential relationships between them, which leads to the joint modeling of both tasks~\cite{hirvonen2015classification,adavanne2018sound}. %In~\cite{adavanne2018sound}, a convolutional neural network (CNN) with discrete  directions-of-arrival (DoAs) as localization targets is used as a multilabel-multiclass  classification problem. On the other hand, ~\cite{hirvonen2015classification} used a regression based approach for localization using a CNN based model with two output branches, one for classification and the other for localization.

% With the progress in 
% deep learning the sound event localization and classification technologies are often developed using
% Either \ac{DOAE} or \ac{SEC} requires large-scale dataset, which is the basis for 
A large-scale database is required for developing deep learning-based approaches to solve real-world problems~\cite{nakamura2000acoustical,sharma2021nhss}.
% Apart  from  developing innovative technology, a proper database is the basis for analyzing and solving real-world problems in a particular area~\cite{nakamura2000acoustical,sharma2021nhss}.
Several  public databases provide resources to explore solutions for \ac{SLC}.
% Several databases exist for \ac{SLC}
% There are several databases for sound localization and sound event classification,
% which provide resources to develop independent technologies for the both the tasks. 
For the \ac{DOAE} sub-task, the SSLR database~\cite{he2018deep} was collected by a 4-channel microphone array mounted on top of a Pepper robot's head. The recorded sounds are transmitted from either a loudspeaker or human subjects.
The RSL database~\cite{sheelvant2019rsl2019} includes  22.6 hours of data, which 
was recorded using a 4-channel microphone array with the sound played over a loudspeaker from 72 different \ac{DoA}s.
For the \ac{SEC} sub-task, the RWCP sound scene database~\cite{nakamura2000acoustical} serves as a long standing corpus, which contains a large number of isolated speech and non-speech environmental sounds.  
% Besides, the DCASE2017~\cite{mesaros2019sound} is another widely used  \ac{SEC} database which includes multiple overlapping sound events.% belong to multiple classes.
% Another weakly labeled database with overlapping events is introduced in DCASE2017~\cite{mesaros2019sound}.
To promote the research and development in \ac{SLC}, a series of DCASE challenges are organized every year, where consequent different databases with varying recording sound sources and acoustic scenarios, are released~\cite{cao2019polyphonic,politis2020dataset}.

% To promote the research and development of sound event localization and classification methods the series of DCASE Challenge is organized every year and consequently different versions of sound event localization and classification databases are released~\cite{politis2020dataset,cao2019polyphonic}. The DCASE 2019~\cite{cao2019polyphonic} challenge presents  a  challenging multi-room reverberant dataset  with varying numbers of overlapping sound events. In addition to this DCASE 2020 challenge dataset~\cite{politis2020dataset} introduces moving sources for about half of the active events. 
  
%This sound event classification and localization challenge provide two datasets referred to as TAU Spatial Sound Events 2019 - Ambisonic (FOA) and TAU Spatial Sound Events 2019 - Microphone Array (MIC), of identical sound scenes with the only difference in the format of the audio. The FOA dataset provides four-channel First-Order Ambisonic recordings while the MIC dataset provides four-channel directional microphone recordings from a tetrahedral array configuration.

Looking into prospects and progress in the field of \ac{SLC} technology, a new publicly available database with different sound events is of great importance. A fairly designed database with different sound event classes recorded from several \ac{DoA}s would help us to solve many real-world SLC problems either by signal processing based analysis, formulating new features or by developing neural networks that learn from a large dataset. 
In this paper, we propose a sufficiently large database, referred to as SLoClas, which differs from the existing databases in terms of sound classes and recording set-up. The SLoClas database is recorded by a 4-channel microphone array, including 10 environmental sound classes omitting from 72 different \ac{DoA}s. We consider realistic sound event classes that include sounds corresponding to bells, bottles, buzzer, cymbals, horn, metal, particle, phone, ring and whistle, each having around 100 examples. The proposed SLC database would provide a common platform to carry on research in DOAE and SEC as two independent tasks, along with the unified SLC task. 
To facilitate the study of noise robustness,  6 types of noise signals are recorded at 4 \ac{DoA}s. Additionally, we collect the Time Stretched Pulse (TSP) signals played from all the positions of the loudspeaker, which can be used to calculate the impulse response of the environment~\cite{suzuki1995optimum}. Besides, we present a baseline framework on SLC using the proposed  SLoClas database that may serve as a reference system for prospective research. We publicly release this database and the source code of the reference SLC system for research purpose~\footnote{\url{https://bidishasharma.github.io/SLoClass/}}.
% Although the proposed SLoClas corpus is well motivated from earlier studies~\cite{cao2019polyphonic,politis2020dataset}, the sound classes, recording set-up and \ac{DoA}s are significantly different from the existing databases.
% As a first step towards development of a real-time robotic audition, our current database is collected in a quiet environment and using single source at a time. 
% We refer the proposed database as SLoClas corpus.

This paper is organized as follows. In \Sec~\ref{sec:datacollection}, we specify the details of the SLoClas database. To provide a comparative analysis for future technologies, we propose a baseline system, where the input acoustic features and the network architecture are discussed in \Sec~\ref{sec:study}.  The implementation details and the experimental results are provided in \Sec~\ref{sec:experiments}.
Finally, we conclude in \Sec~\ref{sec:conclusion}.
% We release this database publicly to provide a common platform to develop technologies for joint sound source localization and classification. We also provide neural network based baseline systems using the proposed SLoClas corpus for comparative analysis of the future technologies.  In this work, we present the details of the effort to collect SLoClas and effectiveness and reference systems for unified sound localization and classification framework.

% \qian{highlight the design of a database, the need of a database, concept, specification, later a case study as a reference benchmark, use our system as a reference design}

\vspace{-0.2cm}
\section{SL\MakeLowercase{o}C\MakeLowercase{las} Database}\label{sec:datacollection}

In this section, we discuss the SLoClas database design and specify
% and specifications of data collection equipment along with
the collection details.

\vspace{-0.3cm}
\subsection{Database design}~\label{databasedesign}
% In this work, 
We use RWCP sound scene classification corpus~\cite{nakamura2000acoustical} as the source database, which consists of three broad categories of environmental sounds, including  crash sounds of wood, plastic and ceramics, sounds occurred when humans operate on things like spray, saw, claps, coins, books, pipes, telephones, toys, etc. The RWCP database consists of 90 sound classes, where each class comprises of around 100 samples. We note that this database was collected in an anechoic room by B\&K 4134 microphone and DAT recorder in 48kHz 16bit sampling with a SNR level around 40-50 dB.

\begin{figure}[tb]
\begin{center} 
\includegraphics[width=\columnwidth]{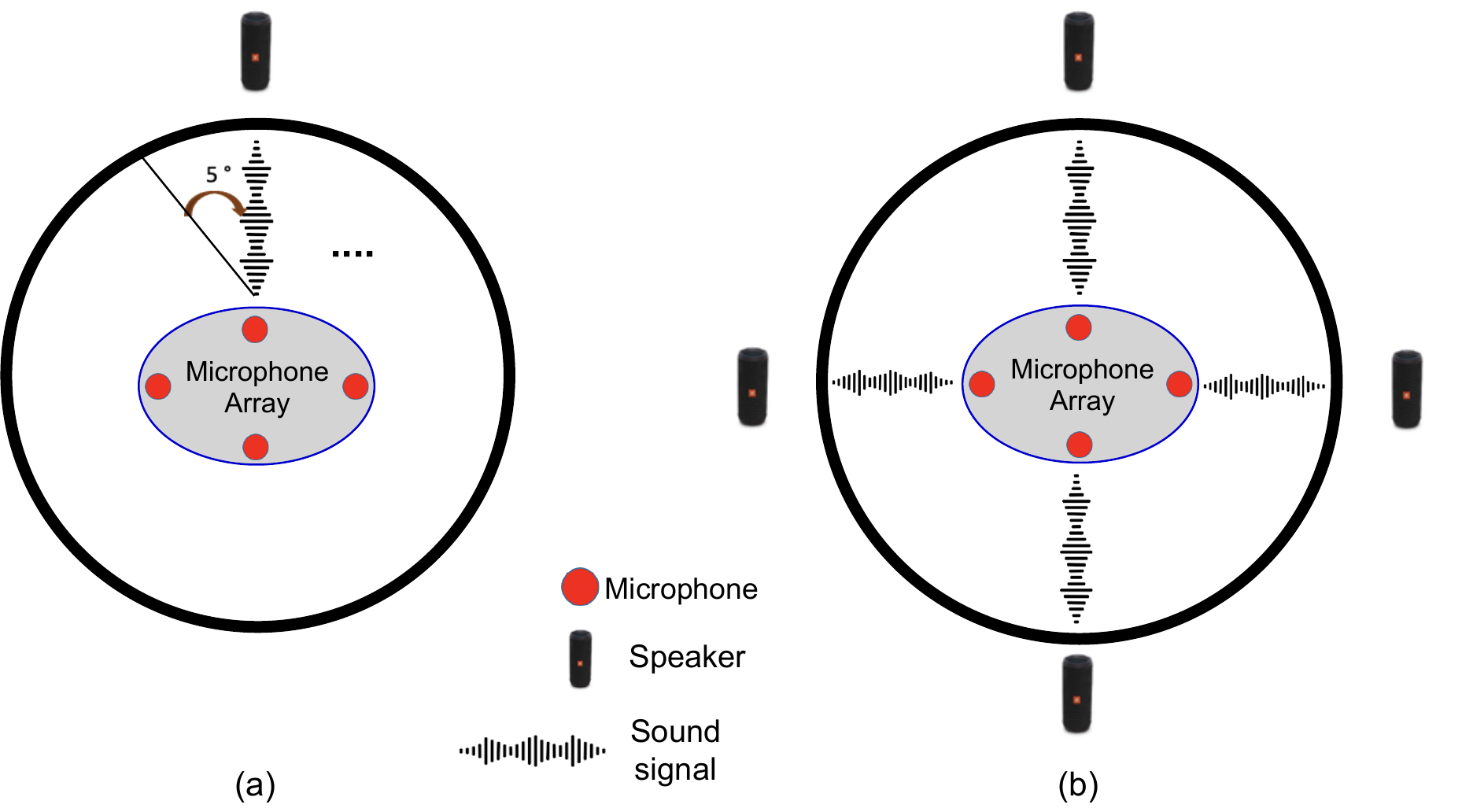}
\end{center}
\vspace{-0.4cm}
\caption{Visualization of the SLoClas database recording setup when (a) record the sound data, (b) record the noise data.}
\label{fig:Sound_Classification_Localisation}
\vspace{-0.3cm}
\end{figure}

We use 10 sound classes from the RWCP corpus (each includes around 100 samples) for the SLoClas corpus collection, and achieve a total of 988 samples. We concatenate these 988 sample sound signals by inserting silence in between, along with the TSP signal combined into a single full length audio file to be played from each position of the loudspeaker.  The microphone array is positioned at the room center and the speaker is shifted by varying the angle from 1$^\circ$ to 360$^\circ$, at 5$^\circ$ interval from 1.5 metre distance as shown in Figure~\ref{fig:Sound_Classification_Localisation}(a). This results in (988$\times$ 72) 71,136 recorded sounds equivalent to 23.27 hours of data.

Besides the aforementioned data, we also record 6 types of outdoor noise (crowd noise, machine noise, traffic and car noise, animal sound, water sound, wind sound) from the PNL 100 Non-speech Sounds database~\cite{hu2010tandem}. The noise data was recorded in a similar way at 4 \ac{DoA}s, keeping the 1.5 metre speaker-array distance as shown in Figure~\ref{fig:Sound_Classification_Localisation}(b). We record 64 noise sounds of 6 classes from each \ac{DoA}, resulting in a total of (64$\times$ 4) 256 noise audio samples.

\vspace{-0.3cm}
\subsection{Recording set-up and equipment}
We use a partially reverberation free professional recording studio of dimension 4 m$\times$5 m. The aforementioned source data is played from a JBL Charge 4 Bluetooth Speaker as discussed in Section~\ref{databasedesign}. The loudspeaker volume is kept at half of the highest loudness setting. We use Audacity audio interface~\cite{Audacity} connected to the microphone array (ReSpeaker Mic Array v2.0) through a laptop, to capture the audio. The ReSpeaker Mic Array contains 4 high  performance  digital microphones,  with  48  kHz  maximum  sampling  rate. ReSpeaker Mic Array v2.0 supports USB Audio Class 1.0 (UAC 1.0), which is directly connected to the laptop. To avoid any obstruction in the sound propagation path, we mount the microphone array and loudspeaker at the same vertical height using stands and the laptop is operated from ground with minimal noise. We use measuring scale and digital angle finder to mark different \ac{DoA}s on a 1.5 metre circle on the ground, according to which the loudspeaker location is varied. The marks are at every 5$^\circ$ azimuth angle ranging from 1$^\circ$ to 360$^\circ$. 
%~\footnote{\url{ https://wiki.seeedstudio.com/ReSpeaker_Mic_Array_v2.0/}}
The noise data is recorded using the same equipment setup from  0\degree, 90\degree, 180\degree, 270\degree  and maintaining the distance of 1.5 meters between the loudspeaker and the microphone array.
% distance between the microphone array and the loudspeaker. 
All audio files are recorded in $f_s$=48kHz sampling rate. 

% \ac{DoA}
\begin{table} [t]
\small
%\vspace{-0.3cm} 
\caption{\label{Sound_class_recording} {{Details of the SLoClas 2021 database
in  recording set-up, sound data, and noise data
(\# indicates number).}}}
\vspace{.3cm} 
%\centerline{
%\scalebox{0.64}{
%\centerline{
% \resizebox{8.5cm}{!}{
\begin{tabular}{p{1.98cm} p{5.75cm}}
\toprule
\midrule
 {\bf Attribute} &  {\bf Details}\\
\midrule
\multicolumn{2}{l}{ \textbf{1. Recording set-up:} } \\
 $f_s$ / Bit rate &  {48 kHz} / 16\\
%  {Bit rate} &  {16} \\
 Sound distance  & 1.5 m (to the array center) \\
  &  {(a) ReSpeaker Mic Array v2.0} \\
  &  {(b) JBL Bluetooth Speaker} \\
Devices &  {(c) two 50-cm stainless steel measuring scales} \\
  &  (d) Digital angle finder {(0.5m/18-inch stainless steel ruler)} \\ \midrule
  \multicolumn{2}{l}{ \textbf{2. Sound data:} } \\
 {\# Sound classes} &  {10 (bells, bottles, buzzer, cymbals, horn, metal, particle, phone, ring, whistle)}\\
 {\# Samples } & {988 per {DoA}}\\
% \hline
 {\ac{DoA} range} &  {1$^\circ$-360$^\circ$ (5$^\circ$ interval)} \\
% \hline
 {\# \ac{DoA} } &  {72} \\
% \hline
 {\# Total samples} &  {71,136} \\
% \hline
 {Total duration} &  {23.27 hours} \\ \midrule
\multicolumn{2}{l}{ \textbf{3. Noise data:} } \\
 {\# Noise classes} &  6 (crowd noise, machine noise, traffic and car noise, animal sound, water sound, wind sound) \\
 {\# Samples} &  64 per DoA \\
DoA range &  {1$^\circ$-360$^\circ$ (90$^\circ$ interval)} \\
 {\# DOA} &  {4} \\
 {\# Total samples} &  {256} \\
 {Total duration} & {13 minutes} \\
 \midrule \bottomrule
\end{tabular}
\vspace{-.4cm}
\end{table} 

% \begin{table} [t]
% \caption{\label{Noise_recording} {{Details of noise recording in SLC 2021 database.}}}
% \vspace{.3cm} 
% \resizebox{8.8cm}{!}{
% \begin{tabular}{|c c|}
% \hline
% \multicolumn{1}{|c|}{\bf Attribute} & \multicolumn{1}{|c|}{\bf Details}\\
% \hline
% \hline
% \multicolumn{1}{|c|}{\#Noise classes} & \multicolumn{1}{|c|}{6}\\
% \hline
% \multirow{2}{*}{Noise classes} & \multicolumn{1}{|c|}{crowd noise, machine noise, traffic and car noise,}\\
% \multicolumn{1}{|c|}{} & \multicolumn{1}{|c|}{animal sound, water sound, wind sound}\\
% \hline
% \multicolumn{1}{|c|}{\#Examples per angle} & \multicolumn{1}{|c|}{64}\\
% \hline
% \multicolumn{1}{|c|}{Range of DOA i.e. angle (degree)} & \multicolumn{1}{|c|}{0-360 (90 degree interval)} \\
% \hline
% \multicolumn{1}{|c|}{\#DOA i.e. angle (degree)} & \multicolumn{1}{|c|}{4} \\
% \hline
% \multicolumn{1}{|c|}{\#Total examples} & \multicolumn{1}{|c|}{256} \\
% \hline
% \multicolumn{1}{|c|}{Total duration} & \multicolumn{1}{|c|}{13 minutes} \\
% \hline
% \hline
% \end{tabular}
% }
% \vspace{-.2cm}
% \end{table} 

\vspace{-.3cm}
\subsection{Post-processing}
% As discussed in Section~\ref{database_design}, we record 988 sound samples, along with TSP signal concatenated into a full length audio file, which played placing the loudspeaker at each angle from 1.5 metre distance.  
For the collected data, we manually annotate the  starting point of each full-length recorded audio file with respect to the source audio in Audacity. Using this manually marked starting label, we synchronize the source (loudspeaker) signal and the 4-channel recorded audio signals.
Considering the start of the audio as an anchor point, we segment all the sample sounds with energy based evidence~\cite{sharma2017vowel,sarma2015exploration,sharma2016sonority} and manual observation. In this way, we achieve 988 segmented audio files and a TSP signal for each DOA angle. We perform a similar processing for the noise data. 
Details of the proposed SLoClas database are depicted in Table~\ref{Sound_class_recording}.
\vspace{-0.2cm}
\section{Reference system on SLoClas database}\label{sec:study}
In this section, we present
% the details of the benchmark system for \ac{SLC} using the proposed SLoClas database. We refer 
a reference system, namely \ac{SLCnet}, to jointly optimize the  \ac{SEC} and \ac{DOAE} objectives for sound localization and classification.
The network uses the sequential hand-crafted acoustic features (i.e. \ac{GCC-PHAT} and \ac{MFCC}) as the inputs and produce a \ac{DoA} label and a sound class label for each audio segment.
% The \ac{SEC} and D\ac{DOAE} objectives jointly through a neural network architecture.
% The \ac{DoA} label and sound class label are predicted at each audio segment.
% We train the \ac{SEC} and DOA estimation \ac{DOAE} objectives jointly through a neural network architecture.

% \ac{SSL}, which identifies the location of sounding object, has played an significant role in many applications, such as robotic listening~\cite{wang2021gcc,qian2021multi}, speech enhancement~\cite{zhang2021deep}, and speaker diarization~\cite{tao2021is}.
% Deep learning-based \ac{DOAE} methods have the advantages of good
% generalization under different levels of reverberation and noise~\cite{cao2019polyphonic}.
% Deep networks are designed to learn the mappings between input acoustic features and the sound \ac{DoA} where experiments demonstrate the promising results.

% In real-world applications, a sound event is always transmitted
% in one or several directions. Given this fact, it is reasonable to combine \ac{SEC} and localization into an unified system and it is worthwhile to study the effects and potential connections between the two tasks.
% These capabilities are essential for robots to separate and identify the sounding object and eventually decide whether reacts to the event or not.
\vspace{-.2cm}
\subsection{Features}
% In this section, we demonstrate the acoustic features employed for \ac{DOAE} and \ac{SEC}.
\vspace{-.2cm}
\subsubsection{GCC-PHAT}
The \ac{GCC-PHAT} is widely used to calculate the TDOA between any two microphones in an  array~\cite{he2018deep,pan2020multitones}. 
We adopt it as the audio feature~\cite{knapp1976generalized} due to its robustness in the noisy and reverberant environment \cite{florencio2008does} and the fewer tunable parameters than the other counterparts \eg~Short Time Fourier Transform (STFT) \cite{chakrabarty2019multi}.
Let ${\bf S}_{p_1}$ and ${\bf S}_{p_2}$ be the Fourier transforms of audio sequence at $p_1$ and $p_2$ channels of the $p$-indexed microphone pair ($p\in [1,...,P]$), respectively. We compute the GCC-PHAT features with different delay lags $\tau$ as:
\begin{equation}\label{eq:gccphat}
    {\bf g}^p(\tau) = \sum_{k}\mathcal{R}\left(\frac{{\bf S}_{p_1}[k]({\bf S}_{p_2}[k])^{*}}{|{\bf S}_{p_1}[k]({\bf S}_{p_2}[k])^{*}|} e^{j \frac{2\pi k}{N} \tau}\right)
\end{equation}
where, $*$ denotes the complex conjugate operation, $\mathcal{R}$ denotes the real part of complex number and $N$ denotes the FFT length. Here, the delay lag $\tau$ between  two signals arrived is reflected in the steering vector  $e^{j \frac{2\pi k}{N} \tau}$ in \eq~\ref{eq:gccphat}. 
\vspace{-.2cm}
\subsubsection{MFCC}
%The GCC-PHAT contains the time-delay information between a microphone pair which is efficient for \ac{DOAE}~\cite{firstgcc}. 
The GCC-PHAT feature is not optimal for \ac{SEC} since it equally sums over all frequency bins while disregarding the sparsity of audio signals in frequency domain where the randomly distributed noise may be stronger than the original audio signal~\cite{firstgcc}. To explore the frequency domain information, we incorporate the \ac{MFCC} features from the 4-channel microphone array, and denote the feature as $\mathbf{M}^m$ where, $m\in [1,...,M]$ is the microphone index.
% \begin{equation}\label{eq:MFCC}
%     \mathbf{M}^m=
% \end{equation}
% where $m\in [1,...,M]$ is the microphone index.

% \qian{to be filled the MFCC equation}
\vspace{-0.2cm}
\subsection{Network architecture}
\begin{figure}[tb]
\begin{center} 
\includegraphics[width=0.8\columnwidth]{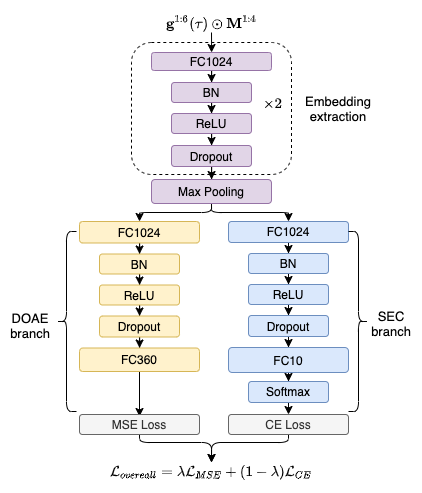}
\end{center}
\vspace{-.6cm}
  \caption{The proposed \ac{SLCnet} architecture for SLC, with 3 main parts: (1) the general embedding extraction block (purple), (2) the \ac{DOAE} branch, (3) the \ac{SEC} branch. We use the weighted \ac{MSE} and \ac{CE} as the final loss. The max pooling is operating along the time dimension (FC: fully-connected layer; BN: batch normalization; $\odot$ indicates concatenation).}
  \label{fig:network}
  \vspace{-.2cm}
\end{figure}

\begin{figure}[tb]
\begin{center} 
\includegraphics[width=\columnwidth]{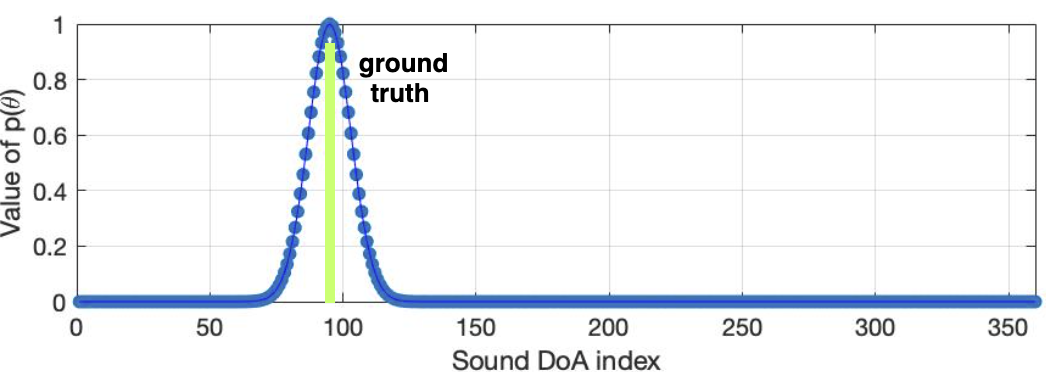}
\end{center}
\vspace{-0.5cm}
  \caption{The sound \ac{DoA} likelihood-based encoding $p(\theta)$ (\eq~\ref{eq:likelihood}) that centralized at the ground truth $\dot{\theta}=95^\circ$.}
  \label{fig:likelihood}
\end{figure}

The network architecture is shown in \fig~\ref{fig:network}, which consists of three parts: (1) the embedding extraction (purple), (2) the \ac{SEC} branch (blue), and (3) the \ac{DOAE} block (yellow). Details of each block are described in the next subsections. For simplicity, we eliminate the time index subscript.
\vspace{-.2cm}

\subsubsection{Embedding extraction}

We concatenate the 6-pair \ac{GCC-PHAT} features and the 4-channel \ac{MFCC} features  as the network inputs. 
We use two stacked \ac{FC} layers, followed by \ac{BN}, ReLU activation, and dropout operation, to generate the latent feature embeddings.
A max pooling operation is applied along the time dimension. Then these extracted embeddings are fed into the  \ac{DOAE} and \ac{SEC} branches.

\vspace{-0.2cm}
\subsubsection{DoA estimation}
We use the likelihood-based coding  \cite{he2018deep} to represent the sound location posterior in each direction. Specifically, each element of the encoded 360-dimensional vector $p(\theta)$ is assigned to a particular azimuth direction $\theta\in [1, 360^\circ]$.  The value of $p(\theta)$ follows a Gaussian distribution that maximizes at the ground truth \ac{DoA}:
\begin{equation}\label{eq:likelihood}
    p(\theta)= exp \left ( -\frac{|\theta-\dot{\theta}|^2}{\sigma^2} \right )
\end{equation}
where, $\dot{\theta}$ is the ground truth, $\sigma$ is a pre-defined constant related to the width of the Gaussian function.  For intuitive understanding, in \fig~\ref{fig:likelihood} we illustrate an example of $p(\theta)$ (blue dotted curve) centralized at $\dot{\theta}=95^\circ$ (vertical green bar).

We predict the sound \ac{DoA} through a deep-learning network, formulated as:
\begin{equation}\label{eq:DoAmodel}
   \hat{p}(\theta)= \mathcal{F}_{\theta}( {\bf g}^{1:M}, \mathbf{M}^{1:M} | \Phi)
\end{equation}
where, $\Phi$ are the trainable parameters, $\mathcal{F}_{\theta}$ is the proposed  \ac{DOAE} network consists of stacked \ac{FC} layers, \ac{BN}, and ReLU activation function.
The $\mathcal{F}_{\theta}$ network corresponds to the purple, green and yellow blocks in \fig~\ref{fig:network}, which outputs the 360-dim \ac{DoA} posterior. 

Finally, we decode the sound \ac{DoA} estimate by finding the maximum of the output:
\begin{equation}\label{eq:decode}
    \hat{\theta}= \text{argmax}_{\forall \theta} \ \hat{p}(\theta)
\end{equation}

Instead of \ac{CE} loss, we adopt the \ac{MSE} loss to measure the similarity between $p(\theta)$ and $\hat{p}(\theta)$, formulated as:
\begin{equation}\label{eq:MSE}
    \mathcal{L}_{MSE}=||p(\theta)-\hat{p}(\theta)||^2_2
\end{equation}
where, $||\cdot||_2$ represents $l$-$2$ distance.

\subsubsection{Event classification}
% For sound event classification, 
We use one-hot encoding for the \ac{SEC} task:
\begin{equation}\label{eq:eventclass}
    y({\dot{c}}) = {\bf 1}_{c=\dot{c}}
\end{equation}
where, $c=1,...,C$ is the event class label with $C$ as the total class number and $\dot{c}$ as the ground truth label.

We predict the event class through the \ac{SEC} network:
% , formulated as:
\begin{equation}\label{eq:eventmodel}
   \hat{y}(c)= \mathcal{F}_{e}( {\bf g}^{1:M}, \mathbf{M}^{1:M} | \Psi)
\end{equation}
where, $\Psi$ are the trainable parameters, $\mathcal{F}_{e}$ consists of stacked \ac{FC}, \ac{BN}, ReLU activation function, and a softmax layer is applied as the output layer. The \ac{SEC} network includes the purple, green and blue blocks in \fig~\ref{fig:network}, which outputs the probability of each sound event.
 
A \ac{CE} loss is computed  between the \ac{SEC} estimate and the ground truth sound class, formulated as:
\begin{equation}\label{eq:CEloss}
    \mathcal{L}_{CE}=-\text{log} \left ( \frac{ \text{exp}(y(\dot{c}))}{\sum^C_{c=1} \text{exp}(\hat{y}(c))} \right )
\end{equation}
where, $\hat{y}(c)$ is computed from \eq~\ref{eq:eventmodel}.

% \subsubsection{Multi-task training}

\begin{figure}[!tb]% \label{fig:visualization}
\begin{center} 
\subfigure[ DOAE (sound localization)]{ \label{subfig:localization}
\includegraphics[width=\columnwidth]{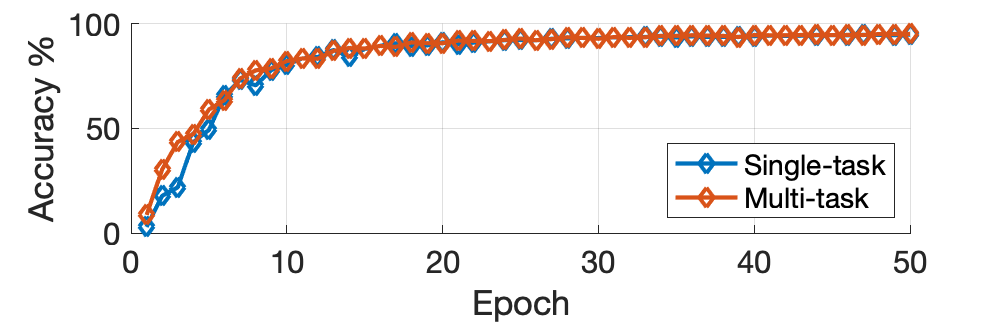}}
\subfigure[SEC (sound event classification)]{\label{subfig:classification}
\includegraphics[width=\columnwidth]{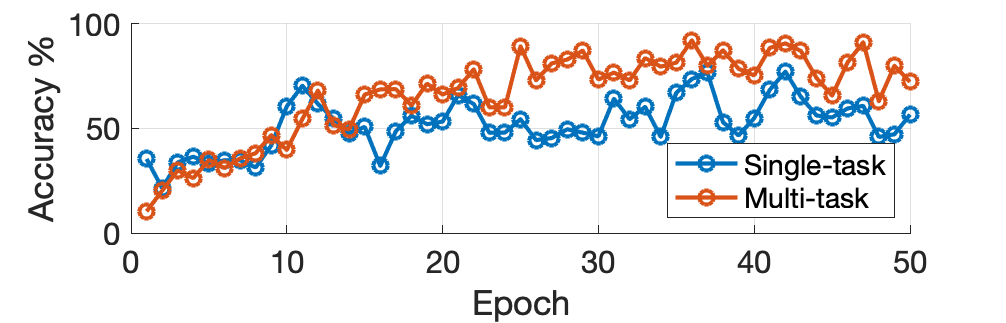}}
\end{center}
\vspace{-0.4cm}
  \caption{The accuracy (\%) variations on the test set over the training epochs (1) for the \ac{DOAE} task, and (2) the \ac{SEC} task. } 
  \label{fig:visualization}
\end{figure}

To optimize a multitask-based network, the following objective function is used for \ac{SLC}:
\begin{equation}\label{eq:loss}
    \mathcal{L}_{overall} = \lambda \mathcal{L}_{MSE} + (1-\lambda) \mathcal{L}_{CE}
\end{equation}
where, $\lambda$ is a pre-defined threshold which indicates the weight of the \ac{DOAE} loss. Moreover, the network corresponds to the single-task \ac{DOAE} when $\lambda=1$ and corresponds to the single-task \ac{SEC} when 
 $\lambda=0$.

\vspace{-0.2cm}
\section{Experiments}\label{sec:experiments}
\vspace{-0.2cm}
\subsection{Experimental setup}\label{sec:intro}
% We train the network for 30 epochs and use report the best result in the test set for this paper.

The \ac{GCC-PHAT} is computed for every 170 ms segments  with the delay lag $\tau \in [-25,25]$, resulting in 51 coefficients for each microphone pair.
Given 6 pairs (select from any two of the 4  microphones) with each pair contributing 51 GCC-PHAT coefficients, we obtain 306 GCC-PHAT coefficients. For \ac{MFCC} features, we average the 4-channel audio into a single-channel audio. We consider a frame size of 20 ms with 50\% overlapping. The MFCC feature include 13-dimensional MFCC coefficients, its delta and delta-delta features, resulting into a 39-dimensional feature set.
% Due to the difference in framesize of MFCC and GCC-PHAT extraction,
We concatenate the MFCC features of 8 successive frames (170 ms) to form a 312-dimensional feature set. In this way, we synchronize the frames of GCC-PHAT and MFCC, and concatenate both features to derive a 618-dimensional feature set to develop the proposed SLC network. % \qian{Details to be filled here} 
 
We use the Adam optimizer \cite{kingma2014adam}. All models are trained for 50 epochs with a batch size of 32  and a learning rate of 0.001. 
% All the neural network training is performed using PyTorch\footnote{https://pytorch.org/}. 
% Since for sound localization, we treat it as a regression task, we use \ac{MSE} instead of \ac{CE} as the loss function. 
Source code in this paper is available~\footnote{\url{https://github.com/catherine-qian/cocosda-SSL.git}}.

%~\footnote{\url{https://bidishasharma.github.io/SLoClass/}}
\vspace{-0.2cm}
\subsection{Evaluation metrics}\label{ssec:metric}
We evaluate the method performance using the \ac{MAE} and the \ac{ACC} metrics.

For \ac{DOAE},  \ac{MAE} is computed as the average angular distance between the sound \ac{DoA} ground truth and estimate:
\begin{equation}
    \text{MAE} (^\circ)= \frac{1}{T} \sum^T_{t=1} |\dot{\theta}_t - \hat{\theta}_t |
\end{equation}
where $\dot{\theta}$ is the \ac{DoA} ground truth, $\hat{\theta}$ is the estimate from \eq~\ref{eq:decode}, $t$ is an audio segment index, and $T$ is the total segment number.

\ac{ACC} of \ac{DOAE} is computed as the percentage of the \ac{DoA} estimate whose error is smaller than the pre-defined threshold:
\begin{equation}\label{eq:ACC}
    \text{ACC}_{\theta} (\%) =\frac{1}{T}\sum^T_{t=1}{\bf 1}_{|\hat{\theta}_t-\dot{\theta}_t|\leq \eta}
\end{equation}
where, ${\bf 1}$ is the indicator function, $\eta$ is the \ac{DoA} error allowance (we set to $5^\circ$ in this paper).

For \ac{SEC}, \ac{ACC} is also used and computed as:
\begin{equation}\label{eq:ACCevent}
    ACC_{e}(\%)=\frac{1}{T}\sum^T_{t=1}{\bf 1}_{\hat{y}_t=\dot{y}_t}
\vspace{-0.2cm}   
\end{equation}
\vspace{-0.6cm} 
%\vspace{-0.4cm}
\subsection{Results}
\begin{table}[!tb]
\caption{The experimental results on SLoClas database (NA: result not applicable). "$\downarrow$" and "$\uparrow$" indicate the lower the better result and the higher the better result, respectively. }\label{tab:result}
 %We use $\lambda=0.99$
\small
\vspace{0.2cm}
\centering
\begin{tabular}{lccc}
\toprule
\midrule
& DOAE & SEC & Multi-task \\  \midrule
% & & & ($\lambda=0.99$)  \\ \hline
\multicolumn{1}{l}{MAE ($^\circ$) $\downarrow$} &    4.68 $^\circ$  & NA    &   4.39$^\circ$    \\ \midrule
\multicolumn{1}{l}{ACC$_{\theta}$ ($\%$) $\uparrow$}  & 94.93\%      & NA    &  95.21\%    \\ \midrule
\multicolumn{1}{l}{ACC$_{e}$ ($\%$) $\uparrow$}  &    NA  &  77.23\%   & 80.01\% \\ \midrule \bottomrule
\end{tabular}
\end{table}

We conduct the experiments on the SLoClas dataset, for which the numerical results are reported in \tab~\ref{tab:result} in terms of MAE, ACC$_{e}$, ACC$_{\theta}$, respectively. 

For the \ac{DOAE} task, we achieve the \ac{MAE} of 4.68$^\circ$ and the accuracy of $94.93\%$. For the \ac{SEC} task, an accuracy of $77.23\%$ is achieved.
For the multi-task training, we report the results with $\lambda=0.99$ (\eq~\ref{eq:loss}) and achieve the  \ac{MAE} of  4.39$^\circ$,  ACC$_{\theta}$ of 95.21\%, and  ACC$_{e}$ of 80.01\%. Compared to the individual task, the multi-task training brings benefits to each of the \ac{DOAE} and \ac{SEC} branch.

\fig~\ref{fig:visualization} illustrates the variations of the accuracy (\%) on the SLC test set over 50 training epochs where the blue colour corresponds to the single-task training (use a single loss function, either \eq~\ref{eq:MSE} or \eq~\ref{eq:CEloss}) and the red colour corresponds to the multi-task training. In particular,  \fig~\ref{subfig:localization} represents  the \ac{DOAE} task and \fig~\ref{subfig:classification} represents  the \ac{SEC} task, respectively.
From \fig~\ref{subfig:localization}, we can visualize that with the increasing training epoch, the \ac{DOAE} performance becomes stable with the resulting \ac{ACC}$_{\theta}>90\%$. Besides, the multi-task training strategy doesn't bring significant improvement for \ac{DOAE}. From \fig~\ref{subfig:classification}, we can observe a higher performance of the red curve than the blue curve, which indicates the benefits to \ac{SEC} from mutli-task training, where the final event classification accuracy equals 80\%.

% \ac{SSL}

% \section{Discussion}\label{sec:intro}
\vspace{-0.2cm}
\section{Conclusion}\label{sec:conclusion}
\vspace{-0.2cm}
In this paper, we present the SLoClas database, to support the study and analysis for joint sound localization and classification. The database contains a total of 23.27 hours of data, including 10 classes of sounds, recorded by a 4-channel  microphone array.
The sound is played over a loudspeaker from different \ac{DoA}s, varying from  1$^\circ$  to 360$^\circ$ at an interval of 5$^\circ$. 
% We consider 10 sound classes with around 100 examples for each class, which are played over the loud-speaker positioned at different angles  from the  microphone array  at  1.5  metre  distance.   
% We  vary  the  \ac{DoA}  from  1$^\circ$  to 360$^\circ$ at an interval of 5$^\circ$.
Additionally, to facilitate the study of noise robustness, we record 6 types of outdoor noise at 4 \ac{DoA}s using the same devices. 
We also propose a reference system (i.e., \ac{SLCnet}) with the experiments conducted on SLoClas corpus. We release the source code and the dataset for use in research purpose. 

\footnotesize
\bibliographystyle{IEEEtran}
\bibliography{main}

\end{document}

%% file: acro.tex
\newacro{SSL}[SSL]{Sound Source Localization}
\newacro{DOAE}[DOAE]{Direction-of-Arrival Estimation}
\newacro{DoA}[DoA]{Direction-of-Arrival}
\newacro{SEC}[SEC]{Sound Event Classification}
\newacro{JECDE}[JECDE]{Joint Event Classification and DoA Estimation}
\newacro{STFT}[STFT]{Short-Time-Fourier-Transform}
\newacro{GCC-PHAT}[GCC-PHAT]{Generalized Cross Correlation with Phase Transform}
\newacro{MSE}[MSE]{Mean Square Error}
\newacro{MAE}[MAE]{Mean Absolute Error}
\newacro{ACC}[ACC]{Accuracy}
\newacro{CE}[CE]{Cross Entropy}
\newacro{MFCC}[MFCC]{Mel Frequency Cepstral Coefficient}
\newacro{FC}[FC]{Fully Connected}
\newacro{BN}[BN]{Batch Normalization}
\newacro{SLCnet}[SLCnet]{Sound Localization and Classification Network}
\newacro{TDOA}[TDoA]{Time Difference-of-Arrival}
\newacro{SRP}[SRP]{Steered Response Power}
\newacro{SLC}[SLC]{Sound Localization and Classification}
\newacro{GMM}[GMM]{Gaussian Mixture Model}
\newacro{SVM}[SVM]{Support Vector Machine}
\newacro{HMM}[HMM]{Hidden Markov Model}